\documentclass[aps,amsfonts,amsmath,prd,preprint,nofootinbib]{revtex4}

\usepackage{bbm}
\usepackage{amsmath}
\usepackage{graphicx}
\usepackage{bm}
\usepackage{amssymb}
\usepackage{latexsym}
\usepackage{graphicx}% Include figure files
\usepackage{dcolumn}% Align table columns on decimal point
\usepackage{bm}% bold math
\usepackage{mathrsfs}

\def\ba{\begin{eqnarray}}
\def\ea{\end{eqnarray}}

\newcommand{\beq}{\begin{equation}}
\newcommand{\eeq}{\end{equation}}

\def\lap{\lower.5ex\hbox{$\; \buildrel < \over \sim \;$}}
\def\gap{\lower.5ex\hbox{$\; \buildrel > \over \sim \;$}}

\begin{document}

\title{Past incompleteness of a bouncing multiverse}
\author{Alexander Vilenkin and Jun Zhang}

\address{Institute of Cosmology, Department of Physics and Astronomy,\\ 
Tufts University, Medford, MA 02155, USA}

\begin{abstract}

According to classical GR, Anti-de Sitter (AdS) bubbles in the multiverse terminate in big crunch singularities. It has been conjectured, however, that the fundamental theory may resolve these singularities and replace them by nonsingular bounces.  This may have important implications for the beginning of the multiverse.  Geodesics in cosmological spacetimes are known to be past-incomplete, as long as the average expansion rate along the geodesic is positive, but it is not clear that the latter condition is satisfied if the geodesic repeatedly passes through crunching AdS bubbles.  We investigate this issue in a simple multiverse model, where the spacetime consists of a patchwork of FRW regions.  The conclusion is that the spacetime is still past-incomplete, even in the presence of AdS bounces.

\end{abstract}
\maketitle

\section{Introduction}

Inflationary cosmology \cite{Guth:1980zm,Linde:1981mu,Albrecht:1982wi} has profound implications for the global structure of spacetime, 
particularly for the beginning and the end of the universe.  Generically, inflation never ends: even though it has ended in our local neighborhood, it still continues in remote regions beyond our horizon \cite{Vilenkin:1983xq,Linde:1986fd}. In this sense, inflation is eternal to the future.  At the same time, the theorem proved in Ref.~\cite{Borde:2001nh} indicates that inflation cannot be eternal to the past.  The theorem states that a past-directed geodesic in any spacetime is incomplete, provided that the expansion rate averaged over the affine parameter along the geodesic is positive: $H_{av}>0$.
%\footnote{This statement applies to all geodesics, except perhaps a set of measure zero.}
The latter condition is expected to hold for past-directed geodesics originating in any inflating region. 
%\footnote{Regions of local contraction could occur in the process of structure formation, but one expects expansion to prevail on average.  In any case, there will be some geodesics that avoid passing through contracting protoclouds.}  
The conclusion is that inflationary spacetimes are necessarily past-incomplete, and thus inflation must have some sort of a beginning.

The spacetime structure of the universe is also influenced by the underlying particle physics model.  Particle theories with extra dimensions, including string theory, predict a vast landscape of vacua with diverse properties \cite{BP,Susskind}.  Combined with inflationary cosmology, this leads to the picture of a multiverse, where bubbles of  different vacua nucleate and expand in the inflating background \cite{CdL}, so the entire landscape is explored.\footnote{Transitions between different vacua can also occur through quantum diffusion \cite{Vilenkin:1983xq,Linde:1986fd}, or through bubble collisions \cite{Jose,Easther:2009ft}.    
For simplicity, here we shall not consider these additional transition mechanisms.} The resulting spacetime structure is schematically illustrated in a causal diagram in Fig.1.  

\begin{figure}[t]
\begin{center}
\includegraphics[width=12cm]{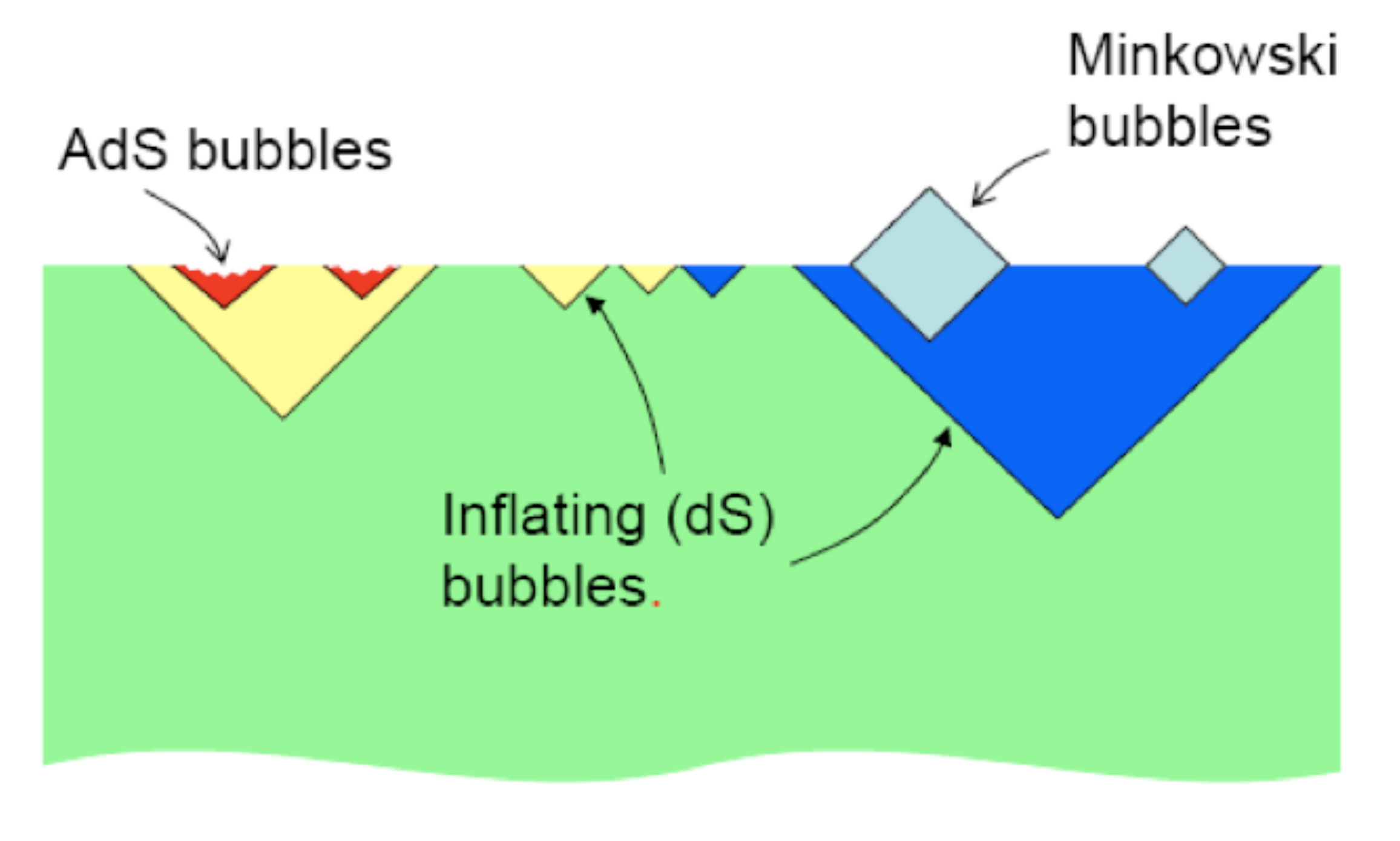}
\caption{A causal diagram of the inflationary multiverse (the standard picture).  The vertical direction represents time, and the horizontal direction is space.  Bubbles of different types, represented by different colors in the diagram, nucleate and expand close to the speed of light.  dS bubbles inflate eternally, and other bubbles nucleate within them.  AdS bubbles collapse to big crunch singularities (indicated by zigzag lines).  Bubbles of stable Minkowski vacua are represented by `hats'.} 
\end{center}
\end{figure}

Disregarding quantum fluctuations, bubble interiors are open FRW universes. If the vacuum inside a bubble has positive energy density, the evolution is asymptotically de Sitter (dS), and the bubble becomes a site of further bubble nucleation.  Negative-energy, anti-de Sitter (AdS) vacua, on the other hand, collapse to a big crunch and develop curvature singularities, which are represented by zig-zag lines in the figure.  The standard assumption is that spacetime terminates at these singularities.\footnote{The particle physics model may also include some stable Minkovski vacua.  Bubbles of such vacua, which are represented by `hats' in the diagram, would form in the multiverse, but they will not be important for our discussion here.}   

It is conceivable, however, that singularities will eventually be resolved in the fundamental theory of Nature, so that AdS crunches will become nonsingular.  The standard description of AdS regions will still be applicable at the initial stages of the collapse, but when the density and/or curvature get sufficiently high, the dynamics would change, resulting in a bounce.  Scenarios of this sort have been discussed in the literature in various contexts 
\cite{Markov0,Markov1,Markov2,Veneziano,Khoury,cyclic,Ashtekar,Peter,Allen,Xue,Creminelli,Lin,Easson,Cai,Brustein}.  Because of the extreme (probably near-Planckian) energy densities reached near the bounce, the crunch regions are likely to be excited above the energy barriers between different vacua, so transitions to other vacua are likely to occur \cite{Piao1,Gupt:2013poa,Garriga:2013cix}.  

The past-incompleteness theorem of Ref.~\cite{Borde:2001nh} does not straightforwardly apply to multiverse models with AdS bounces.  The past of inflating regions now includes not only other inflating regions, but also contracting AdS regions, so it is not obvious that the average expansion rate $H_{av}$ along past-directed geodesics is necessarily positive.  Thus, AdS bounces open an intriguing possibility of a past-eternal universe.  In the present paper, we shall study this possibility by directly calculating the affine length of past-directed null
geodesics.  Our analysis here will be less general than that in Ref.~\cite{Borde:2001nh}.  We shall disregard possible gravitational effect of the bubble walls and inhomogeneities caused by quantum fluctuations inside the bubbles.  Thus, we shall assume that bubble interiors have open FRW geometry.  Our conclusion is that the multiverse spacetime is still past-incomplete, even in the presence of AdS bounces.

The paper is organized as follows. In the next Section we specify our model assumptions.  The matching conditions for the affine parameter of null geodesics at bubble boundaries are derived in Section III.  In Section IV, these conditions are used to calculate the affine length of past-directed null geodesics in a simple model with one dS and one AdS vacuum.  This analysis is extended to a multi-vacuum landscape in Section V and to more general FRW components (not pure dS or AdS) in Section VI.  Finally, our conclusions are summarized and discussed in Section VII.

\section{The model}

We shall approximate the multiverse spacetime by a patchwork of dS and AdS regions, which are matched together according to the following rules.  

1. dS and AdS bubbles are bounded by the future lightcones of their nucleation points (we shall refer to them as bubble cones).  We disregard the gravitational effect of the bubble walls (which would otherwise perturb the spacetime outside of the bubble cones).

2. Bubble interiors are described by open FRW metrics,
\beq
ds^2 = dt^2 - a^2(t)(d\chi^2 + \sinh^2\chi d\Omega^2) .
\label{metric}
\eeq
For a dS bubble the scale factor is
\beq
a(t) = H_{dS}^{-1} \sinh (H_{dS}t) ,
\label{dS}
\eeq
and for an AdS bubble it is
\beq
a(t) = H_{AdS}^{-1} \sin (H_{AdS}t) .
\label{AdS}
\eeq
The surface $t=0$ is the bubble cone, and the condition $da/dt(0) = 1$ guarantees that the geometry remains smooth on that surface.  The constants $H_{dS}$ and $H_{AdS}$ are generally different for different bubbles.

3. Bubbles nucleate in dS regions at a constant rate $\Gamma$ per spacetime volume, which depends both on the type of bubble and on the parent dS vacuum.  (The actual values of the nucleation rates will not be important in what follows.)  We disregard the possibility of bubble nucleation in AdS regions.

4. To simplify the analysis, we shall assume that the dS geometry of the parent bubble can be continued into the daughter bubble up to some small time $t=\delta$, where $t$ is the time coordinate in the daughter bubble metric.  That is, we shall assume that $a(t)$ in the daughter bubble is given by Eq.~(\ref{dS}) with the same $H_{dS}$ as in the parent bubble for $0\leq t\leq \delta$.  Alternatively, this part of the daughter bubble spacetime can be described by extending the coordinate system of the parent bubble into this spacetime region.  
This assumption is not very restrictive, since $\delta$ can be made arbitrarily small, and $a(t) \approx t$ at small $t$ for all values of $H_{dS}$ or $H_{AdS}$.

5. An AdS bubble with a scale factor (\ref{AdS}) would collapse to a big crunch singularity at $t_c = \pi/H_{AdS}$, but we shall assume that instead the collapse terminates at $t_b = t_c - a_b$ and is followed by a bounce, which is generally accompanied by a transition to another vacuum.  (Thus the AdS form of the scale factor (\ref{AdS}) applies for $\delta\leq t\leq t_b$.)  We assume that $a_b \ll H_{AdS}^{-1}$, so the scale factor at the bounce is $a(t_b)\approx a_b$.  At later times, the scale factor is
\beq
a(t) = H_{dS}^{-1} \sinh [H_{dS}(t-t_b +a_b)] 
\eeq
if the new vacuum is dS and
\beq
a(t) = {H'}_{ads}^{-1} \sin [{H'}_{ads}(t-t_b +a_b)] 
\eeq
if it is AdS.  

6. We disregard possible effects of bubble collisions.  

Although not very realistic, these assumptions capture the main features of a multiverse spacetime with AdS bounces.  We shall analyze the past-completeness of such spacetimes in Sections III and IV and will extend the analysis to a more general class of models in Sections V and VI.

\section{Affine Parameter of Null Geodesics}\label{AP}

A spacetime is said to be past-incomplete if there is a null (or timelike) geodesic maximally extended to the past, which has a finite affine length. Hence, a direct way of checking geodesic completeness is to calculate the affine length of null geodesics.  In an open FRW universe (\ref{metric}), a radial null geodesic obeys
\beq
a(t)\frac{d\chi}{dt} = 1,
\label{achit}
\eeq
and the affine parameter $\lambda$ can be found from 
\beq\label{inta}
d\lambda = {N} a\left( t \right) dt,
\eeq
where ${N}$ is a normalization constant.  In a multiverse spacetime, we will have to deal with two complications.  First, the geodesic will pass through a number of different bubbles.  Within each bubble the spacetime is FRW, but we need to determine how the normalization factor ${N}$ changes from one bubble to the next.  And second, the propagation will not generally be radial.  We shall first derive the matching condition for the affine parameter in the simpler case of radial geodesics and then extend it to the general case.

\subsection{Radial geodesics}

\begin{figure}[t]
\begin{center}
\includegraphics[width=12cm]{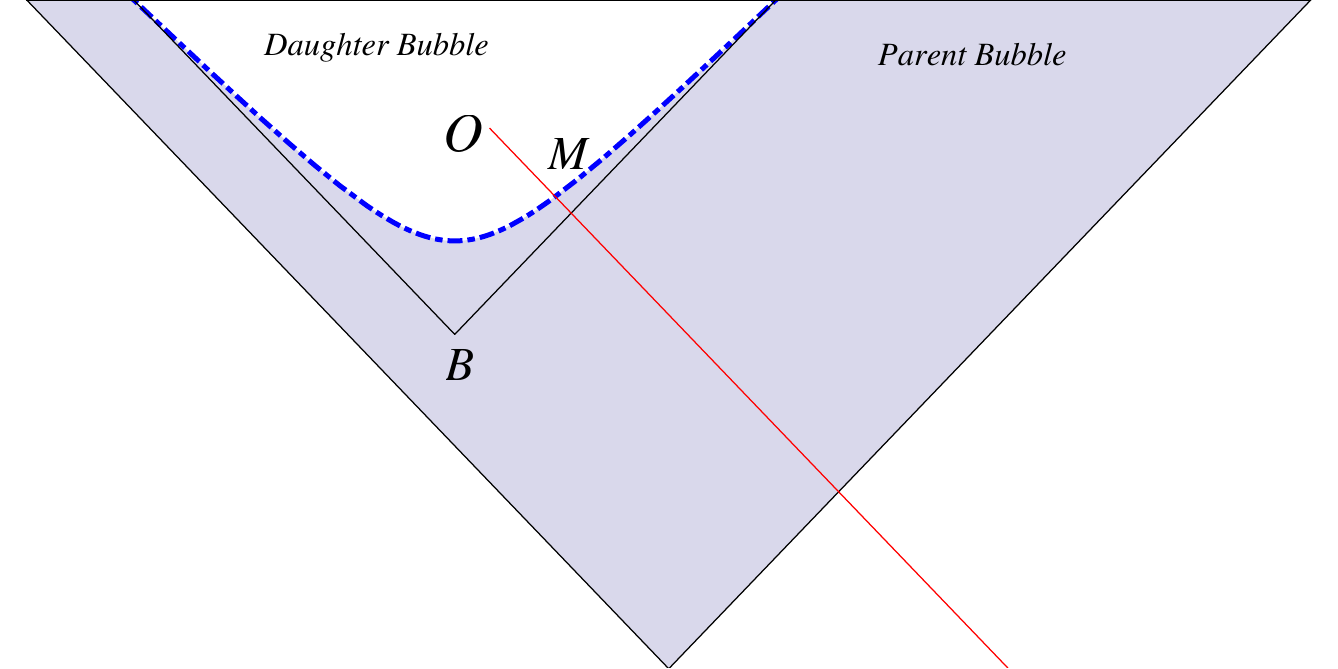}
\caption{A daughter bubble nucleates at point $B$ inside its parent bubble. The red line is a past-directed null geodesic. The shaded area is the overlap region, which can be covered by the coordinates of both parent and daughter bubbles.}\label{figmat}
\end{center}
\end{figure}

Suppose a daughter bubble covered by FRW coordinates $(t', \chi')$ with a scale factor $a'(t')$ nucleates at point $B$ in a parent bubble covered by coordinates $(t, \chi)$ with a (generally different) scale factor $a(t)$.
We consider a radial null geodesic propagating in the $\chi$ direction (with fixed angular coordinates) in the parent bubble and then crossing, also radially, into the daughter bubble, as shown in Fig. \ref{figmat}.  The affine parameter on this geodesic is normalized by ${N}$ in the parent bubble, and we want to find the normalization factor ${N}'$ in the daughter bubble.  

As the geodesic propagates through the daughter bubble, it intersects the congruence of comoving timelike geodesics of the bubble, originating at the nucleation center $B$.  We shall derive the matching condition for the affine parameter by requiring the continuity of the invariant scalar product $u_\mu dx^\mu /d\lambda$, where $u^\mu$ is the 4-velocity of the comoving test particles and $dx^\mu/d\lambda$ is the tangent vector of the null geodesic.  In the coordinates of the daughter bubble, this is 
\ba\label{udx1}
u'_\mu\frac{dx'^{\mu}\left(\lambda\right)}{d\lambda} = \frac{1}{N'}\frac{1}{a'\left(t'\right)} .
\ea
%where $u'^\mu$ is the two-velocity (corresponding to the four-velocity in the 3+1D spacetime) of these test particles, and $x'^{\mu}\left(\lambda\right)$ is the coordinates of the null observer as a function of $\lambda$.

According to the assumption 4 of Section II, the daughter bubble spacetime up to the time $t' = \delta$ can be covered by both coordinate systems, $(t, \chi)$ and $(t', \chi')$.  We shall refer to the corresponding spacetime region as the overlap region.  In the parent bubble's coordinates, timelike geodesics originating at $B$ are no longer comoving; they satisfy
\ba
a^2 \frac{d\chi}{ds} = C,
\ea
where $s$ is the proper time along the geodesic and $C$ is an integration constant. The 4-velocity $u^\mu$ is given by
\ba
\left(\frac{dt}{ds}, \frac{d\chi}{ds} \right)=\left(\sqrt{1+\frac{C^2}{a^2}},\frac{C}{a^2}\right).
\ea
For the null geodesic, Eqs.~(\ref{achit}),(\ref{inta}) give
\ba
\frac{dx^{\mu}}{d\lambda}=\frac1N \left(\frac1a,-\frac1{a^2}\right).
\ea
Thus the scalar product (\ref{udx1}) is 
\ba\label{udx2}
u_\mu\frac{dx^{\mu}}{d\lambda} = \frac{1}{Na^2\left(t\right)}\left(\sqrt{a^2\left(t\right)+C^2}+C\right).
\ea

In order for the affine parameter to continue smoothly from parent to daughter bubble, we require that the expressions in Eqs.~(\ref{udx1}) and (\ref{udx2}) should be equal to one another.  This condition can be imposed at any point $M$ lying on the null geodesic in the overlap region.  We shall choose it to be the point where the geodesic crosses the surface $t' = \delta$.  Let $(t_M, \chi_M)$ be the coordinates of this point and $(t_B, \chi_B)$ the coordinates of the nucleation point $B$ in the parent bubble.  Then we find
\ba\label{nn2}
\frac{N}{N'} =\frac{a'\left(\delta \right)}{a^2\left(t_M\right)} \left(\sqrt{a^2\left(t_M\right)+C^2}+C\right)
\ea
and
\ba\label{snm}
\delta = \int_{t_B}^{t_M}\frac{a\left(t\right)}{\sqrt{a^2\left(t\right)+C^2}}dt .
\ea

\subsection{Non-radial geodesics}

Let us now see how the matching condition (\ref{nn2}) is modified if the null geodesic is not assumed to be radial.  We can choose the matching point $M$ as the origin of spatial coordinates in both parent and daughter bubbles.  This can always be done, since the equal time surfaces in both bubbles are homogeneous and isotropic hyperbolic spaces.  With this choice, both null and timelike geodesics crossing at $M$ will become radial.
%\footnote{Technically, we choose the comoving timelike geodesic that connect two points, the matching point and the point where the parent bubble nucleates, as the origin of the spacial coordinates. We always have this freedom, because inside bubble, the constant time surface is a homogenous and boundless 3D hypersurface.} 
Then the equations of the preceding subsection would still apply everywhere expect at the origin, where the spherical coordinate system cannot be used.  We shall therefore calculate the scalar product using a local Cartesian coordinate system.

In Cartesian coordinates, we can write the 4-velocity $u^\mu$ and the tangent vector of the null geodesic $dx^\mu/d\lambda$ as
\ba
\left(\frac{dt}{ds}, \frac{d\vec{x}}{ds} \right)=\left(\sqrt{1+\frac{C^2}{a^2}},\frac{C}{a^2} \vec{n}\right) ,
\ea
%and for the null geodesics $x^\mu\left(\lambda\right)$, we have
\ba
\frac{dx^{\mu}}{d\lambda}=\frac1N \left(\frac1a,-\frac1{a^2}\vec{\nu}\right) ,
\ea
where $\vec{n}$ and $\vec{\nu}$ are unit 3-vectors in the directions of timelike and null geodesics, respectively. 
The scalar product in the coordinates of parent bubble then becomes
\ba\label{udx3}
u_\mu\frac{dx^{\mu}}{d\lambda} = \frac{1}{Na^2\left(t\right)}\left(\sqrt{a^2\left(t\right)+C^2}+C\cos\theta\right),
\ea
where $\cos\theta \equiv \vec{n}\cdot\vec{\nu}$. The calculation in the daughter bubble is unchanged, and we conclude that the matching condition (\ref{nn2}) is replaced by
\ba\label{nn4}
\frac{N}{N'} =\frac{a'\left(\delta \right)}{a^2\left(t_M\right)} \left(\sqrt{a^2\left(t_M\right)+C^2}+C\cos\theta  \right) .
\ea

\subsection{Small $\delta$ limit}

With $a(t)$ from Eq.~(\ref{dS}) and $H_{dS}=H_p$, corresponding to the parent bubble, Eq.~(\ref{snm}) gives
\ba\label{s3}
\cosh\left(H_{p}t_M\right) = \cosh\left(H_{p}t_B\right)\cosh\left(H_{p}\delta\right)+ \sinh\left(H_{p}\delta\right)\sqrt{\sinh^2\left(H_{p}t_B\right)+H_{p}^2C^2}.
\ea
As we noted in Sec. II, the matching time $\delta$ can be chosen arbitrarily small.  Hence, for any $t_M$ and $t_B$, $t_M > t_B$, we can choose $\delta$ such that 
$\delta \ll t_M - t_B$.  Then it follows from Eq.~(\ref{snm}) that $C \gg a\left(t_M\right)$, and Eq. (\ref{s3}) becomes
\ba
\cosh\left(H_{p}t_M\right) \simeq \cosh{\left(H_{p}t_B\right)} + H_{p}^2\delta C.
\ea
Substituting this in (\ref{nn4}) and using $a'(\delta)\approx\delta$, we obtain
\beq
\frac{N}{N'} \simeq \frac{\delta C}{a^2\left(t_M\right)} (1+\cos\theta) \simeq \frac{\cosh\left(H_{p}t_M\right)-\cosh\left(H_{p}t_B\right)}{\sinh^2\left(H_{p}t_M\right)} (1+\cos\theta).
\label{matching}
\eeq

Note that this final form of the matching condition is independent of the value of $\delta$.  In what follows we shall adopt the limit $\delta\to 0$.  That is, we shall disregard the overlap regions, assuming that the transitions from parent to daughter bubble geometry occur directly on the bubble cones.

\section{Two-vacuum Model}

We can now use the matching condition (\ref{matching}) to investigate the geodesic completeness of the multiverse spacetime.  We shall start with a simple two-vacuum toy model and then generalize to a multi-vacuum landscape in the following section.

%and show that, even with considering the contractions in AdS bubbles, the spacetime is still past-incomplete. 

\begin{figure}[t]
\begin{center}
\includegraphics[width=11cm]{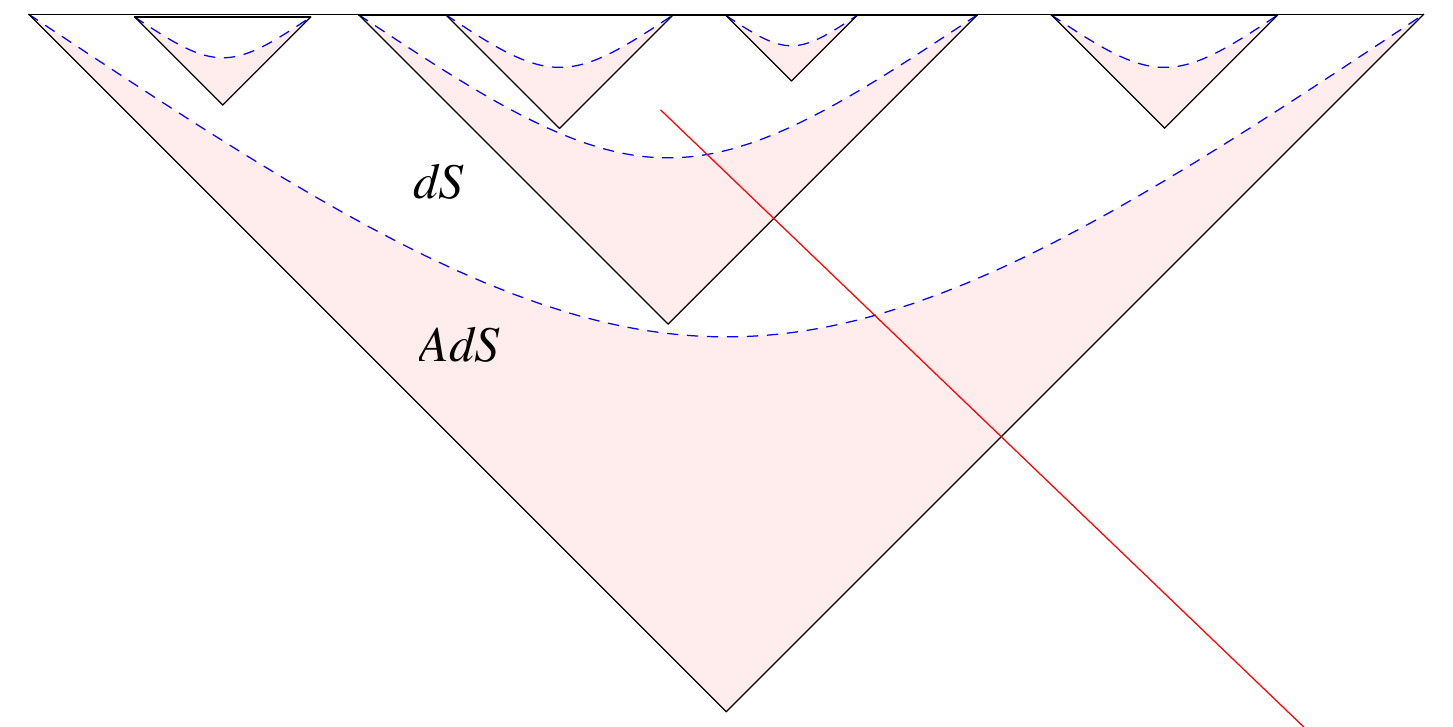}
\caption{A two-vacuum toy model.  Pink regions are in the AdS vacuum, and white regions are in the dS vacuum.  The blue dashed lines mark the bounces where AdS transits to dS.}
\label{toymodel}
\end{center}
\end{figure}

The landscape of our toy model consists of one dS and one AdS vacuum.  As shown in Fig. \ref{toymodel}, transitions from AdS to dS occur at nonsingular bounces, while transitions from dS to AdS occur through bubble nucleation. 
%(more precisely, it occurs a small time $\delta$ after bubble nucleation).  
A null geodesic in this spacetime will pass through an infinite succession of dS and AdS regions.  We shall refer to a part of the geodesic contained within a single bubble as one cycle.  Each cycle starts when the geodesic enters a bubble and ends when it enters a daughter bubble nucleated inside that bubble.
%However, as we mentioned in Sec. \ref{AP}, the coordinates of parent but can also cover a little region inside the daughter bubbles, so the scale factor observed by this null geodesic in each cycle can be written as
The corresponding scale factor can be written as
\ba
a\left(\tau\right) = \left\{ 
  \begin{array}{l l}
    H_{AdS}^{-1}\sin\left(H_{AdS} \tau \right) & 0\le \tau \le \tau_b\\
    H_{dS}^{-1}\sinh{\left[H_{dS}\left(\tau-\tau_b+a_b\right)\right]} & \tau_b \le \tau \le \tau_{M}
  \end{array} \right. 
\label{cycle}
\ea
Here, $\tau_b$ is the time of the bounce, $a _b = \frac{\pi}{H_{AdS}}-\tau_b$ is the scale factor at the bounce (we assume, as before, that $a_b \ll H_{AdS}^{-1}$), and $\tau_M$ is the end of the cycle (which is different for different cycles). 

For the convenience of the following calculation, we shall introduce a new time coordinate $t$,
\beq
t \equiv \tau-\tau_b+a_b ,
\eeq
so that the bounce occurs at $t=a_b$ and the cycle evolution in Eq.~(\ref{cycle}) is replaced by
\ba
a\left(t\right) = \left\{ 
  \begin{array}{l l}
    -H_{AdS}^{-1}\sin\left[H_{AdS}\left( t-2a_b \right)\right] & t_i \le t \le a_b\\
  H_{dS}^{-1} \sinh{\left(H_{dS}t\right)} & a_b \le t \le t_M .
  \end{array} \right.. 
\ea
Here, $t_i = -\frac{\pi}{H_{AdS}} +2a_b$ and $t_M = \tau_M -\tau_b +a_b$ are respectively the beginning and the end of the cycle is the new time coordinate.  The motivation for this choice of time coordinate is that the scale factor in the de Sitter region now has the form (\ref{dS}) that we assumed in Sec.~III, so the matching condition (\ref{matching}) can be directly applied.

The affine length of a null geodesic in one cycle is $\lambda = N\widetilde\lambda$, where the unnormalized affine length is given by
%integrated from the matching point in $\left(j+1\right)^{th}$ cycle to the matching point in $j^{th}$ cycle,
\ba
\widetilde\lambda = \int_{t_i}^{t_M} a\left(t\right) dt 
\simeq\frac{2}{H_{AdS}^2}+\frac{1}{H_{dS}^2}\left[\cosh\left(H_{dS}t_M\right)-1\right].
\label{tildelambda}
\ea
%We can find that the evolution is symmetric near the bounce, and after bounce the space becomes exponential expansion asymptotically.\footnote{In this case, the bounce is curvature dominated. The bounce could also be kinetic energy dominated where $a\left(t\right) \sim \left|t\right|^{1/3}$ near the bounce. But it will not make too much difference in the problem of our interests, if we assume bubble nucleation only starts after vacuum energy dominates.}
As we follow a past-directed null geodesic back in time, we label the cycle in which the geodesic originates by $j=0$, and the subsequent cycles by $j = -1, -2, -3...$.  Choosing the normalization $N_0 = 1$ in the original cycle, we can express the total affine length of the geodesic from cycle $j=-1$ to $-\infty$ as  
%The non-normalized affine length $\widetilde\lambda$ in $j^{th}$ cycle is integrated from the matching point in $\left(j+1\right)^{th}$ cycle to the matching point in $j^{th}$ cycle,
%\ba
%\widetilde\lambda_j \equiv \int_s^{t_M} a\left(t\right) dt \simeq\frac{2}{H_{AdS}^2}+\frac{1}{H_{dS}^2}\left[\cosh\left(H_{dS}t_M\right)-1\right].
%\ea
%Then if we normalize the affine length in the $0^{th}$ bubble, the affine length of this geodesic is normalized as
\ba
%\label{totlam}
\lambda &=& \sum_{\substack{j = -1} }^{-\infty} {N_{j}}\widetilde\lambda_{j} \nonumber\\
&=& \sum_{\substack{j = -1} }^{-\infty}\left(
\prod_{\substack{k = -1} }^{k = j}\frac{N_{k}}{N_{k+1}}\right) \widetilde\lambda_{j} ,
\label{lambda}
\ea 
where ${\widetilde\lambda}_j$ are given by Eq.~(\ref{tildelambda}) with values of $t_M=t_{Mj}$ which are generally different for each cycle.  Introducing the notation
\ba
B_j \equiv \frac{N_{j}}{N_{j+1}}\widetilde\lambda_j ,
\label{Bj}
\ea
we can rewrite this as
\beq
\lambda = B_{-1} + \sum_{\substack{j = -2} }^{-\infty}\left(\prod_{\substack{k = -1} }^{k = j+1}\frac{N_{k}}{N_{k+1}}\right)B_j .
\label{lambda2}
\eeq

From the matching condition (\ref{matching}), we can write
\ba
\frac{N_j}{N_{j+1}} < 2 \frac{\cosh\left(H_{dS}t_{Mj}\right)-1}{\sinh^2\left(H_{dS}t_{Mj}\right)}  < \frac{2} {\cosh\left(H_{dS}t_{Mj}\right)+1} \equiv \alpha_j < 1.
\label{Nbound}
\ea
We also find that
\ba
B_j &\simeq& \frac{N_j}{N_{j+1}}\left[\frac{2}{H_{AdS}^2}-\frac{1}{H_{dS}^2}\right]+\frac{2}{H_{dS}^2} \beta_j,
\ea
where,
\beq
\beta_j \equiv \frac{\cosh\left(H_{dS}t_{Mj}\right)-\cosh\left(H_{dS}t_{Nj}\right)}{\sinh^2\left(H_{dS}t_{Mj}\right)}\cosh\left(H_{dS}t_{Mj}\right).
\eeq
Denoting $X = e^{2H_{dS}t_{Mj}} > 1$ and $Y = e^{H_{dS}\left(t_{Mj}-t_{Nj}\right)} \ge 1$, we have
\ba
\beta_j &=&\left(\frac{X+1}{X-1}\right)^2-\frac{X+1}{\left(X-1\right)^2}\left(\frac{X}{Y}+Y\right)\nonumber\\ 
&\le& \frac{\left(X+1\right)^2-\sqrt{X}\left(X+1\right)}{\left(X-1\right)^2} < 1,
\ea
Hence, $B_j$ is bounded by
\ba
B_j < \frac{2}{H_{AdS}^2}+\frac{2}{H_{dS}^2} \equiv B_{max}
\ea
Then it follows from Eq.~(\ref{lambda2}) that the total affine length should satisfy
\beq
\lambda < B_{max} S ,
\eeq
where
\beq
S = 1 + \alpha_{-1} + \alpha_{-1}\alpha_{-2} + ... .
\label{S}
\eeq

The quantities $\alpha_j$ depend on the times $t_{Mj}$ that the geodesic spends in the $j$-th bubble.  They are generally different for different geodesics and can be thought of as independent random variables, taken from some distribution ${\cal P}(t_M)$.  For completeness, we calculated this distribution in Appendix A, even though  
its form is not important for our analysis here.

The past completeness of the multiverse spacetime depends on the convergence of the sum ${S}$ in  Eq.~(\ref{S}).  This stochastic sum will be different for different geodesics, but its average value can be easily calculated.  From Eq.~(\ref{S}) we can write
\beq
  S = 1 + \alpha_{-1} (1+\alpha_{-2}+\alpha_{-2}\alpha_{-3} + ...) \equiv 1 +\alpha_{-1} S' ,
\eeq   
where $S' = 1+\alpha_{-2}+\alpha_{-2}\alpha_{-3} + ...$.  Since all $\alpha_j$ are independent, this implies the following relation for the average of $S$:
\beq
\langle S\rangle = 1 + \langle\alpha\rangle \langle S\rangle ,
\label{Salpha}
\eeq
and thus
\beq
\langle S\rangle = \frac{1}{1-\langle\alpha\rangle}.
\label{Sav}
\eeq
Since $\alpha_j <1$, we must have $\langle\alpha\rangle <1$.  (The average value $\langle\alpha\rangle$ is calculated in Appendix A in terms of $H_{dS}$ and bubble nucleation rate $\Gamma$.)  Hence, $0<\langle S\rangle<\infty$ and
\ba
\langle\lambda\rangle < 2\left(\frac{1}{H_{AdS}^2}+\frac{1}{H_{dS}^2}\right) \langle S\rangle <\infty.
\ea 
This shows that the affine length of past-directed null geodesics is finite, except perhaps for a set of measure zero,  and thus the spacetime in this two-vacuum model is geodesically past-incomplete.

\section{Multi-vacuum landscape}

%In our toy two-vacuum model, all bubbles had the same structure, consisting of an AdS region, followed by a dS region, with the same values of $H_{AdS}$, $H_{dS}$, $a_b$ and $\delta$

We now consider a landscape including a number of dS and AdS vacua.  We shall assume that any dS vacuum can make transitions to other (dS or AdS) vacua through bubble nucleation.  The transitions rates to different vacua are generally different, and some of them may be zero.  We also assume that any AdS vacuum transits to some other vacuum through a nonsingular bounce.  

As before, we consider a past-directed null geodesic and divide it into cycles, with each cycle contained within a single bubble.  For a dS bubble, the whole cycle is in the same dS vacuum.  For an AdS bubble, the cycle starts in the AdS vacuum and transits to another vacuum after the bounce.  
%(We assume that the new vacuum is different from the initial one.)  
If the new vacuum is AdS, this is followed by another bounce transition.  So an AdS bubble will generally visit a number of AdS vacua, until it finally transits to a dS vacuum.  The cycle ends when the null geodesic enters a daughter bubble in that dS vacuum.

For simplicity we shall assume that AdS bounces are deterministic -- that is, the sequence of vacua visited during a cycle which starts with an AdS vacuum is fully determined by that vacuum.\footnote{Vacuum transitions at AdS bounces may have a stochastic character, due to amplification of quantum fluctuations in the tunneling scalar field by tachyonic instability or by parametric resonance.  However, it has been shown in \cite{Garriga:2013cix} that these mechanisms are typically much less efficient in AdS bounces than in models of slow roll inflation.  Hence the assumption of deterministic transitions may not be very unrealistic.}  Then, in a finite landscape, there is a finite number of possible cycles; we shall label them by letters $a,b,...$ from the beginning of the Latin alphabet.\footnote{We assume that there are no closed AdS loops, resulting in periodic infinite sequences of AdS vacua.  If such sequences did exist, they would include past-eternal geodesics.  However, the second law of thermodynamics requires that the entropy density should be maximized in such an oscillating spacetime.  This implies that observers do not exist in such regions of the multiverse, and we are justified to ignore them.}  Note that these labels are different from $j=1,2,...$, numbering the cycles as they are encountered along the geodesic.  

The total affine length of a past-directed null geodesic can be expressed as in Eqs.~(\ref{lambda2}),(\ref{Bj}),
where $\widetilde\lambda_j$ is the unnormalized affine length in the $j^{th}$ cycle. 
%We already showed that $N_k/N_{k+1} < 1$. 
For dS bubbles we have
\ba
B_j \simeq \frac{2}{H_j^2} \frac{\cosh\left(H_{j}t_M\right)-\cosh\left(H_{j}t_B\right)}{\sinh^2\left(H_{j}t_M\right)}\left[\cosh\left(H_{j}t_M\right)-1\right] \le \frac{2}{H_j^2} ,
\ea
and for AdS bubbles we have 
\ba
B_j \le \frac{2}{H_{dS}^2} + \sum_{AdS} \frac{2}{H_{AdS}^2} ,
\label{BjAdS}
\ea
where the summation is over all AdS vacua encountered in the bubble before it transits to the final dS vacuum with a Hubble constant $H_{dS}$. 

The quantity on the right-hand side of (\ref{BjAdS}) depends only on the type of cycle and can take only a finite set of values.  As before, we can define $B_{max}$ to be the maximum of this quantity over all cycles, and it follows from Eq.~(\ref{lambda2}) that 
\beq
\lambda < B_{max} \left[ 1 + \sum_{\substack{j = -2} }^{-\infty}\left(\prod_{\substack{k = -1} }^{k = j+1}\frac{N_{k}}{N_{k+1}}\right)\right] .
\label{lambda3}
\eeq
 
Let us now consider the ratios $N_j/N_{j+1}$ which determine the change in normalization of the affine parameter from one cycle to the next.  Suppose the cycles $j+1$ and $j$ are of type $a$ and $b$, respectively.  
From Eq.~(\ref{Nbound}) we see that the upper bound on $N_j/N_{j+1}$ depends on the properties of both $a$ and $b$ cycles (through the Hubble parameter $H_j$ and through the time $t_{Mj}$, which depends on the nucleation rate of $a$ bubbles in $b$). 

As before, we shall be interested in the average affine length of a past-directed null geodesic.  The main differences are that {\it (i)} we now have to average over histories involving different types of cycles and {\it (ii)} 
the average will depend on the type $a$ of the initial cycle where the geodesic starts. Denoting this average by $\langle \lambda_a\rangle$ and following the logic that led us to Eq.(\ref{Salpha}), we obtain the relations
\beq
\langle \lambda_a\rangle < B_{max} \langle S_a\rangle ,
\label{lambdaav}
\eeq
\beq
\langle S_a\rangle = 1 + \sum_{\substack{b} } \kappa_{ab}  \langle S_b\rangle .
\label{Skappa}
\eeq
Here, $\langle S_a\rangle$ is the average of the expression in the square brackets in Eq.~(\ref{lambda3}) with the sequence of cycles starting with a cycle of type $a$ at $j=0$ and $\kappa_{ab}$ are given by
\beq
\kappa_{ab}\equiv p_{ab} \langle \alpha_{ab} \rangle ,
\label{kappaab}
\eeq
where $p_{ab}$ is the probability that a cycle of type $a$ is preceded by (or followed by in the backward time direction) a cycle of type $b$, 
\beq
\sum_{\substack{b} } p_{ab} = 1 ,
\label{pab}
\eeq
and 
\beq
\langle \alpha_{ab} \rangle = \langle\frac{2} {\cosh\left(H_{b}t_{ab}\right)+1} \rangle < 1.
\label{alphaab}
\eeq
The time $t_{ab}$ is the time spent by the geodesic in a bubble of type $b$, until it hits a daughter bubble of type $a$ nucleated in $b$ ($b$ follows $a$ in the backward time direction).  The average value $\langle \alpha_{ab} \rangle$ is calculated in Appendix A.

Eq.~(\ref{Skappa}) can be rewritten as
\beq
\sum_{\substack{b} } M_{ab}\langle S_b \rangle = I_a ,
\label{MS}
\eeq
where 
\beq
M_{ab} = \delta_{ab} - \kappa_{ab}
\label{Mab}
\eeq
and $I_a$ is a column vector with components $I_a = \{1,1, ..., 1\}$.  The matrix $M_{ab}$ is a diagonally dominant matrix, which means that it satisfies
\beq
|M_{aa}| > \sum_{\substack{b\neq a} } |M_{ab}|.
\eeq
According to Levy-Desplanques theorem, such matrices are non-singular: $\det M_{ab}\neq 0$.  
It follows that Eq.~(\ref{MS}) has a unique solution with $\langle S_a\rangle <\infty$.  Moreover, all $\langle S_a\rangle$ in this solution satisfy $\langle S_a\rangle > 1$.  To show this, let $S_{min}$ be the smallest of $\langle S_a\rangle$.  Then it follows from Eq.~(\ref{Skappa}) that 
\beq
\langle S_a\rangle > 1 +  S_{min} \sum_{\substack{b} } \kappa_{ab} .
\label{Smin}
\eeq
This inequality should also apply when $\langle S_a\rangle = S_{min}$; then noticing that $\sum_{\substack{b} } \kappa_{ab} < 1$, we see from (\ref{Smin}) that $S_{min}>1$.  (This needed to be checked to make sure that our solutions for $\langle S_a\rangle$ do not include meaningless negative values.)
We thus conclude that the affine length of past-directed null geodesics is finite, except perhaps for a set of measure zero.

\section{More general FRW components}

The model multiverse spacetimes we considered so far consist of a patchwork of dS and AdS regions joined together according to certain rules.  Our analysis, however, can be extended to a wider class of models, where the dS and AdS components are replaced by more general FRW spacetimes.  To simplify the discussion, we shall focus on a two-component model, generalizing the two-vacuum models of Sec.~IV.  The only change we introduce in that model is that the scale factor (\ref{cycle}) is now replaced by a more general form,
\ba
a\left(\tau\right) = \left\{ 
  \begin{array}{l l}
    a_1 \left( \tau \right), & 0\le \tau \le \tau_b\\
    a_2 \left(\tau\right), & \tau_b \le \tau \le \tau_{M} 
  \end{array} \right. 
\label{cycle}
\ea
Here, the function $a_1(\tau) \geq 0$ satisfies the conditions
\beq
a_1(0) = 0, ~~~~ {\dot a}_1(0) = 1 ,
\label{a1cond}
\eeq
it is assumed to reach a maximum value $a_{max}$ somewhere in the middle of its range and to drop to a very small value at $\tau=\tau_b$, 
\beq
a_1(\tau_b) \equiv a_b \ll a_{max}.
\eeq
The conditions (\ref{a1cond}) ensure that the bubble smoothly matches to the background spacetime along the bubble cone.

The function $a_2(\tau)$ describes an expanding FRW universe with a positive energy density.  Then it follows from the Friedmann equation
\beq
{\dot a}^2 = -k +\frac{8\pi G}{3}\rho a^2
\eeq
with $k=-1$ that 
\beq
{\dot a}_2(\tau) > 1.
\label{dota2}
\eeq
$a_2(\tau)$ should also satisfy the continuity condition, $a_2(\tau_b) = a_b$.  As before, we assume that bubbles can only nucleate in the $a_2$ part of spacetime and that the parent spacetime can be continued for a small time $\delta$ into the bubbles.

With $\delta$ sufficiently small, the constant $C$ in Eq.~(\ref{snm}) should satisfy $C\gg a_2(t_M)$; so that
\beq
C\delta \approx \int_{\tau_B}^{\tau_M} a_2(\tau)d\tau , 
\eeq
where $\tau_B > \tau_b$ is the bubble nucleation time. Then the matching condition (\ref{nn4}) yields
\ba
\frac{N}{N'} < \frac{2\int_{\tau_{b}}^{\tau_{M}} a_2\left(\tau\right) d\tau}{a_2^2\left(\tau_{M}\right)} \equiv \alpha(\tau_M) ,
%\le \frac{\int_{\tau_{Sj}}^{\tau_{Mj}} a_j\left(\tau_j\right) d\tau_j}{a_j^2\left(\tau_{Mj}\right)}
\label{Nalpha}
\ea 
and using Eq.~(\ref{dota2}) we can write
\ba
\frac{N}{N'} < \frac{2\int_{\tau_{b}}^{\tau_{M}} a_2\left(\tau\right){\dot a}_2(\tau) d\tau}{a_2^2\left(\tau_{M}\right)} < 1.
%\le \frac{\int_{\tau_{Sj}}^{\tau_{Mj}} a_j\left(\tau_j\right) d\tau_j}{a_j^2\left(\tau_{Mj}\right)}
\ea 

Our proof of past incompleteness in Sec.~IV relied on the inequality $N/N' <1$ and on the fact that the quantity $B_j$ is bounded from above, $B_j < B_{max} < \infty$.  This line of argument, however, cannot be extended to the general case.  From the definition of $B_j$ and Eqs.~(\ref{Nalpha}),(\ref{tildelambda}), we can write
\beq
B_j < 2 \frac{\left( \int_{\tau_{b}}^{\tau_{Mj}} a_2 (\tau) d\tau \right)^2} {a_2^2\left(\tau_{Mj}\right)} .
\label{Bjbound}
\eeq
If the asymptotic form of $a_2(\tau)$ is exponential,
\beq
a_2(\tau) \propto e^{H\tau} ~~~~ (\tau\to\infty) ,
\eeq
then the right-hand side of (\ref{Bjbound}) approaches a finite value $2/H^2$ at $\tau_{Mj} \to\infty$, and $B_j$ is bounded from above.  However, if the expansion is asymptotically power-law,
\beq
a_2(\tau) \propto {\tau}^n ~~~~ (\tau\to\infty) ,
\label{an}
\eeq
then $B_j \propto \tau_{Mj}^2$, so $B_j$ can take arbitrarily large values at large $\tau_{Mj}$.  (Note that for $n>1$ the $a_2$ regions can be inflationry and can contain an infinite number of bubbles.)

We shall therefore take an alternative approach, focusing on the {\it average} affine length of null geodesics from the start.  The total affine length of a geodesic is given by Eq.~(\ref{lambda2}), and we note that the quantity $B_j$ in each term of the sum in this equation depends only on the part of the geodesic in vacuum $j$ and is statistically independent of the factors $N_k/N_{k+1}$ multiplying it in that term.
The average affine length should then satisfy
\beq
\langle\lambda\rangle < \langle B\rangle \langle S\rangle ,
\eeq
where $S$ is the sum defined in Eq.~(\ref{S}).  The average value of this sum is finite and is given by Eq.~(\ref{Sav}).  Thus, the completeness of geodesics depends on whether or not the average
\beq
\langle B\rangle = \int_{\tau_b}^\infty {\cal P}(\tau_M) B(\tau_M) d\tau_M
\label{Bav}
\eeq
is finite.  The asymptotic form of the probability distribution ${\cal P}(\tau_M)$ for a power-law expansion (\ref{an}) is found in Appendix \ref{appB}; for $\tau_M\to\infty$ it is given by
\beq
{\cal P} \left(\tau_M\right) \simeq   \frac{4\pi \Gamma  \tau_M^3}{3\left(n-1\right)^3}\exp\left(-\frac{\pi\Gamma }{3\left(n-1\right)^3}\tau_M^4  \right).
\eeq
With $B(\tau_M)\propto \tau_M^2$, the integral in (\ref{Bav}) is convergent, and thus the null geodesics are past-incomplete, except possibly for a set of measure zero.

%where $\tau_{Bj}$ is when the daughter bubble nucleates and $\tau_{Mj}$ is when the geodesic hits the daughter bubble. So $\tau_{Bj} \ge \tau_{Sj}$. It is obvious that the upper boundary of $N_j/N_{j+1}$ only depends on the evolution after $\tau_{Sj}$. For example, if we ask $a_j\left(\tau_j\right)$ satisfying ${\dot a}_j(\tau_j)\geq 1$ after $\tau_{Sj}$, which is a reasonable requirement for starting nucleation, we can always find $N_j/N_{j+1} \le 1$. Hence the spacetime will be past-incomplete, as long as $\int_{0}^{\tau_{Sj}} a_j\left(\tau_j\right)d\tau_j$ is finite.}
%It can be shown, for example, that in the two-vacuum model of Section IV, the dS scale factor in the second line of (\ref{cycle}) can be replaced by any function $a(\tau)$ satisfying ${\dot a}(\tau_b)\geq 1$ and ${\ddot a} >0$.  

\section{Conclusions}

In this paper we investigated how the past geodesic incompleteness of multiverse spacetimes is affected by AdS bounces.  The criterion of positive average expansion rate derived in \cite{Borde:2001nh} cannot be straightforwardly applied to geodesics traversing many dS and AdS bubbles.  So instead of using that criterion, we calculated the total affine length $\lambda$ of past-directed null geodesics.  We found that $\lambda <\infty$ for all geodesics, except perhaps for a set of measure zero.  Thus, in the class of models that we considered here, the spacetime  is past-incomplete, and the multiverse must have some sort of a beginning.  

This result can be regarded as an extension of the past-incompleteness theorem of Ref.~\cite{Borde:2001nh}, but our analysis here has been less general.  In particular, we disregarded the gravitational effects of bubble walls and of bubble collisions and inhomogeneities that could be generated by quantum fluctuations inside the bubbles.  We assumed also that bubbles can nucleate only in dS regions (or, more generally, in the inflating regions described by the scale factor $a_2(\tau)$ in Sec.~VI).
It would be interesting to investigate possible extensions of our results to a more general class of spacetimes.

%For instance, would it be possible to lift the assumptions of homogeneity and isotropy of the bubble spacetimes?  The following argument suggests that such a generalization should indeed be possible. 

We would like to conclude with the following observations.  The theorem of Ref.~\cite{Borde:2001nh} relates past incompleteness to the average expansion rate $H_{av}$ along a null geodesic.  If $H_{av}>0$, then the geodesic must be past-incomplete.  The quantity $H_{av}$ is calculated for a congruence of timelike geodesics, which does not have to be globally defined: it is enough to define it along the null geodesic of interest.
In our patchwork model, $H_{av}$ is not easy to calculate, since the comoving congruence of geodesics changes discontinuously across the bubble boundaries (or, more precisely, across the bubble cones).  For this reason we used a different criterion of past incompleteness.  But as a matter of principle, it should be possible to smooth the geodesic congruence at bubble crossings.  (Note that comoving geodesics in the parent dS bubble become nearly comoving in the daughter bubble within a few Hubble times after the bubble crossing \cite{Winitzki}.) We can tell, qualitatively, what such a smoothed congruence will look like.  The expansion rate $H$ will be nearly constant and positive in dS regions, it will continue smoothly into AdS regions, remain positive for a while, and then turn negative in the contracting part of the AdS bubble.  As the crunch approaches, $H$ will get large and negative, but then swiftly change to large and positive after the bounce.  The sign of $H_{av}$ depends on whether  expansion or contraction wins on average.  

Our results in this paper suggest that expansion should on average prevail, at least in the models that we considered here.  But suppose for a moment that there is some more general bouncing multiverse model in which $H_{av}<0$. The spacetime in such a model might be past geodesically complete, but then it would have a different problem, of a rather unusual kind.  If $H_{av}<0$, then the same argument that proved incompleteness to the past in Ref.~\cite{Borde:2001nh} would now prove incompleteness {\it to the future}.  This would be a somewhat bizarre and perplexing conclusion.  Future-incomplete geodesics would indicate that the spacetime can be extended beyond what appears to be its future boundary.  But the evolution of our model from given initial conditions is completely specified (at least in a statistical sense) by the field equations, complemented by a semiclassical model of bubble nucleation.  
Since future incompleteness of inflating spacetimes appears rather unlikely, the above argument suggests that our past incompleteness result is more general, extending well beyond the patchwork models for which we proved it here.\footnote{The conclusion of past-incompleteness may be avoided if the dynamics of inflation and bubble nucleation does not extend all the way to $t\to -\infty$.  This kind of picture is adopted in the `emergent universe' scenario, which assumes that an inflating universe emerges from a static or oscillating initial seed \cite{EM2004,Murugan,Mulryne}.  In this case, $H_{av} = 0$ for past-directed geodesics and $H_{av} >0$ for future-directed geodesics, so the spacetime can be complete in both time directions. The problem with this scenario is that the initial seed is generally unstable with respect to particle production and to quantum tunneling \cite{Mithani,Graham,Mithani2}.}

%One way to avoid the conclusion of future-incompleteness is to assume that $H_{av}<0$ at $t\to -\infty$ and 
%$H_{av} >0$ at $t\to +\infty$.  The problem with this picture is that eternal inflation generally exhibits attractor behavior, so the statistical properties of the multiverse do not change with time, except perhaps for a finite period after the beginning -- if there was a beginning.  If the spacetime is complete both to the past and to the future, then we expect $H_{av}$ to be the same at both past and future infinity.  Since future incompleteness of inflating spacetimes appears rather unlikely, the above argument suggests that our past incompleteness result is more general, extending well beyond the patchwork models for which we proved it here.  A possible loophole is that $H_{av}$ might be equal to zero.  It is not clear, however, what mechanism would enforce this condition.  And if it is somehow enforced, a spacetime with a zero average expansion rate is subject to the danger of a thermal death.  It would be interesting to see if these arguments can be made more precise.

\section*{Acknowledgements}

This work was supported in part by the National Science Foundation (grant PHY-1213888) and the Templeton Foundation.  We are grateful to Jaume Garriga and Alan Guth for very useful and stimulating discussions.

\appendix

\section{Probability of hitting a bubble}
\label{appA}

In this appendix, we would like to calculate the average value $\langle \alpha_{ab} \rangle$ (or $\langle \alpha \rangle$ in the two-vacuum model). For the multi-vacuum landscape, we assume bubbles of type $a$ nucleate in bubbles of type $b$ with a constant nucleation rate $\Gamma_{ab}$ per 4-volume. So the total nucleation rate in the type $b$ bubble is $\Gamma_b =  \sum_{\substack{a} } \Gamma_{ab}$. In the case of two-vacuum model, it simply becomes $\Gamma_{ab}=\Gamma_b=\Gamma$.

For a parent vacuum of a given type $b$, the probability of having no bubble nucleation in a 4-volume $\Omega$ is
\ba
P_b \left( \Omega \right) = e^{- \Gamma_{b} \Omega},
\ea
Thus for a null geodesic,  the probability of hitting a type $a$ bubble per unit time after spending time $t_M$ in the type $b$ bubble is given by
\ba
{\cal P}_{ab} \left(t_M\right) = \frac{\Gamma_{ab}}{\Gamma_{b}} \frac{dP_b\left(\Omega_{b}\left(t_M\right)\right)} {dt }  =\Gamma_{ab} \frac{d \Omega_b\left(t_M\right)}{dt_M} \exp\left[ - \Gamma_{b} \Omega_b\left(t_M\right) \right] ,
\ea
where $\Omega_{b}$ is the 4-volume of the past light cone of point $M$ inside the type $b$ bubble. 
%Given the normalization, $\sum_{\substack{a}}\int_{0}^{\infty}{\cal P}_{ab}\left(t\right)dt =1$, we have
%\ba
%{\cal P}_{ab} \left(t_M\right) =\Gamma_{ab} \frac{d \Omega_b\left(t_M\right)}{dt_M} Exp\left[ - \Gamma_{b} \Omega_b\left(t_M\right) \right].
%\ea

\begin{figure}[t]
\begin{center}
\includegraphics[width=11cm]{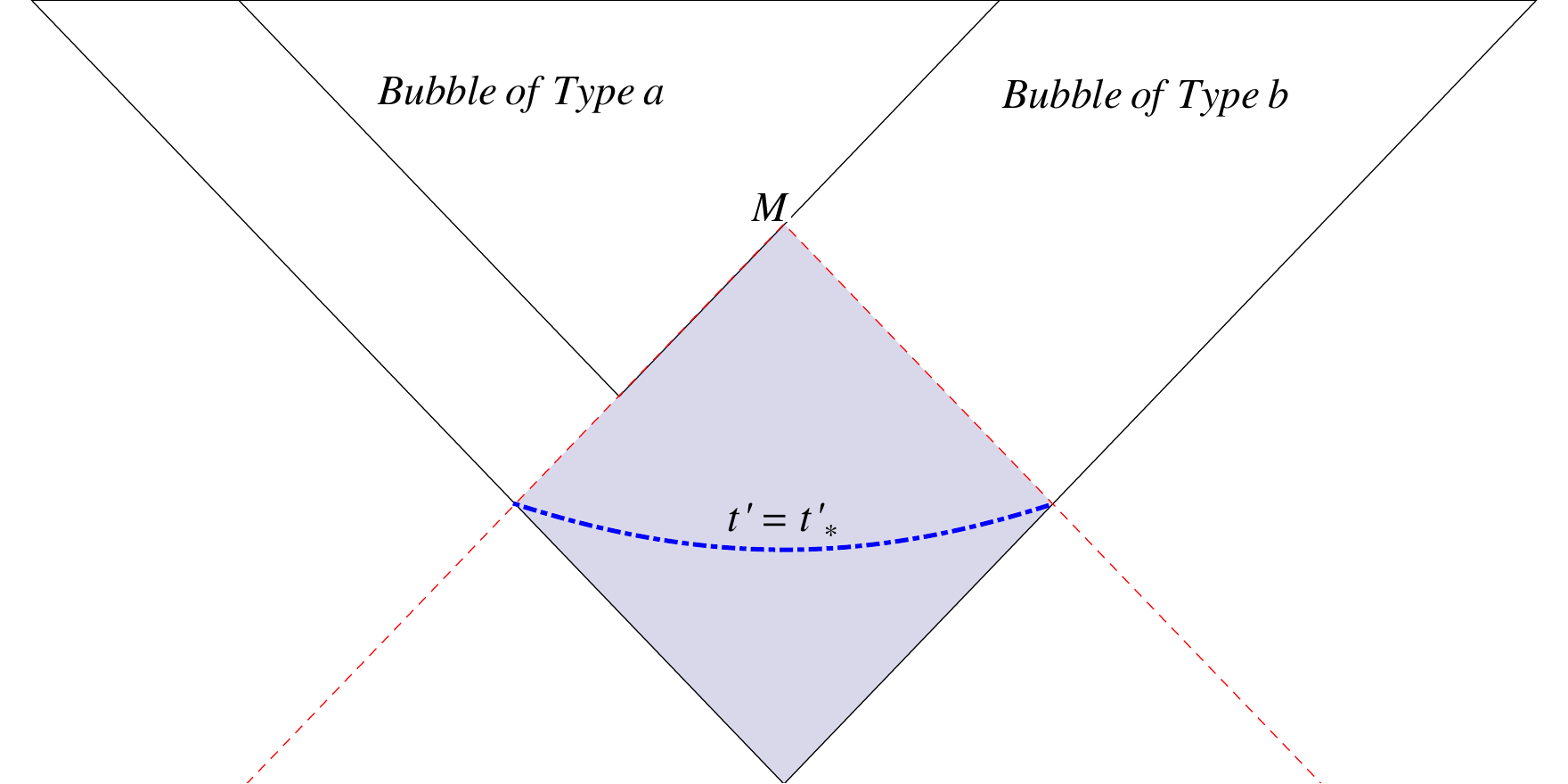}
\caption{A bubble of type $a$ nucleates inside the bubble of type $b$. The red dashed line is the past-light cone of point $M$. The blue dot-dashed line is the surface of $t'=t'_*$. $\Omega_b$ is the 4-volume of the shaded region.}\label{Omega}
\end{center}
\end{figure}

As shown in Fig.\ref{Omega}, the 4-volume $\Omega_b$ is bounded by the bubble cone of type $b$ bubble and by the past-light cone of point $M$. Without loss of generality, we can choose $M$ to be located at $r=0$. Furthermore, to simplify the calculation, we introduce two new coordinates, $(t', r')$, which relate the original coordinates by
\ba
&&e^{H_b t'} = \cosh \left(H_b t\right) + \cosh\left(H_b r\right)\sinh\left(H_b t\right) \nonumber \\
&&e^{H_b t'}r' = \sinh\left(H_b r\right)\sinh\left(H_b t\right).
\ea
So the dS metric in bubble $b$ can be written as
\ba
ds^2 = dt'^2 - e^{2H_b t'}\left(dr'^2+r'^2 d\Omega^2\right),
\ea
and the coordinates of point $M$ are $(t'=t_M, r'=0)$. Using the new coordinates, the past-light cone of point $M$ is given by
\ba
H_b r'_L = e^{-H_b t'} - e^{- H_b t'_M},
\ea
and the bubble cone of the parent bubble $b$ is given by
\ba
H_b r'_W = 1- e^{-H_b t'}.
\ea
Hence, the relevant 4-volume is 
\ba
\Omega_b = \int_0^{t'_*} \frac{4\pi}{3}{r'_W}^3\left(t'\right) e^{3H_bt'} dt' + \int_{t'_*}^{t'_M} \frac{4\pi}{3}{r'_L}^3\left(t'\right) e^{3H_bt'} dt',
\ea
where $t'_*$ is the time when the past-light cone of $M$ and the bubble cone intersect; it can be found from
\ba
2e^{-H_b t'_*} = 1+ e^{-H_b t'_M}.
\ea
Then we find
%\ba
%\Omega_b\left(t'_M\right) = \frac{4\pi}{3H_{b}^3}\left[t'_M - \frac{1}{H_b}\tanh^2\left(\frac{H_bt'_M}{2}\right)\right],
%\ea
%considering $t'_M = t_M$, the 4-volume in the original coordinate system is
\ba\label{apv}
\Omega_b\left(t_M\right) = \frac{4\pi}{3H_{b}^3}\left[t_M - \frac{1}{H_b}\tanh^2\left(\frac{H_bt_M}{2}\right)\right] ,
\ea
where we have used the fact that $t'_M = t_M$.  Note that for $t_M$ much greater than Hubble time, $t_M \gg H_b^{-1}$, $\Omega_b$ can be approximated by a simple formula
\beq
\Omega_b\left(t\right) \approx \frac{4\pi t}{3H_{b}^3} .
\label{Omegaapprox}
\eeq
Then we have
\ba
{\cal P}_{ab} \left(t_M\right) = \frac{4\pi\Gamma_{ab}}{3H_b^3} \exp\left[  -\frac{4\pi\Gamma_{b}}{3H_b^3} t_M \right],
\label{Papprox}
\ea

The distribution ${\cal P}_{ab}(t_M)$ can now be used to calculate the average value
\ba\label{avealpha}
\langle \alpha_{ab} \rangle = \int_{0}^{\infty} \alpha_{ab}\left(t_{M}\right) {\cal P}_{ab}\left(t_{M}\right) d t_{M} ,
\ea
where $\alpha_{ab}(t_M)$ is given by
\beq
\alpha_{ab}(t_M) = \frac{2} {\cosh\left(H_{b}t_M\right)+1} 
\eeq
With the approximate form (\ref{Papprox}) for ${\cal P}_{ab}$, this gives
\ba
\langle \alpha_{ab} \rangle = \frac{\Gamma_{ab}}{\Gamma_b} 2D\left\{1+D\left[\psi\left(\frac{D+1}{2}\right)-\psi\left(\frac{D+2}{2}\right)\right]\right\},
\label{psi}
\ea
where $D\equiv \frac{4\pi\Gamma_b}{3H_b^4}$ and $\psi\left(x\right)$ is the digamma function.  
The approximation (\ref{Omegaapprox}) is expected to be accurate in the limit of low nucleation rate, $D\to 0$.  In this limit, Eq.~(\ref{psi}) reduces to a simple formula
\beq
\langle \alpha_{ab} \rangle \approx \frac{8\pi \Gamma_{ab}}{3H_b^4}.
\eeq

\begin{figure}[t]
\begin{center}
\includegraphics[width=11cm]{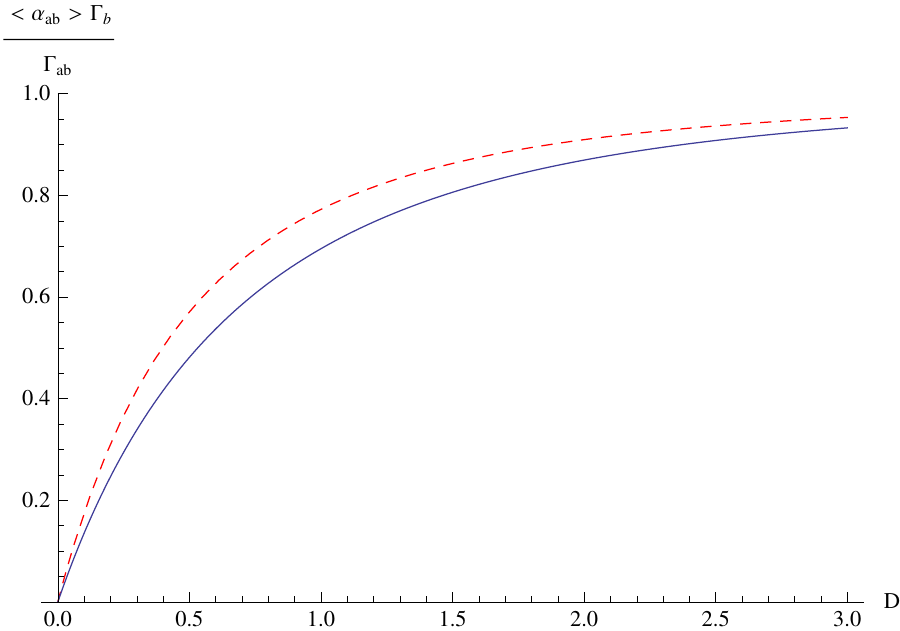}
\caption{The dashed red line is the analytic result calculated with $\Omega_b\left(t\right) = \frac{4\pi t}{3H_{b}^3}$. The solid blue line is calculated with the 4-volume given by Eq. (\ref{apv}).} 
%The numerical integral of Eq.(\ref{avealpha}) is cut off at $H_bt_M = 10^8$, where the curve almost doesn't change with the cutoff.}
\label{alpha}
\end{center}
\end{figure}

We can also calculate $\langle \alpha_{ab} \rangle$ numerically using Eq.(\ref{apv}) for $\Omega_b(t_M)$. 
%The result converges very quickly, since the function we integrated in Eq.(\ref{avealpha}) decreasing exponentially. 
Both analytic and numerical results are shown in Fig.\ref{alpha}.  Vacua with $D>3$ do not exhibit eternal inflation;\footnote{The physical dS volume in a comoving region at $t\to\infty$ is proportional to $\exp[(3-D)H_b t]$, so eternal inflation occurs only for $D<3$.} hence we are only interested in $0 < D <3$.

\section{Probability of hitting a bubble for a power-law expansion}
\label{appB}
In this appendix, we would like to calculate the probability distribution, ${\cal P}\left(\tau_M\right)$, for a power-law expansion described by
\ba
a_2\left(\tau\right)=a_b\left(\frac{\tau}{\tau_b}\right)^n,
\ea 
with $n > 1$. As we already showed in Appendix \ref{appA},
\ba\label{PB}
{\cal P} \left(\tau_M\right) =\Gamma \frac{d \Omega \left(\tau_M\right)}{d\tau_M} \exp\left[ - \Gamma \Omega \left(\tau_M\right) \right] .
\ea
In this case, $\Omega\left(\tau_M\right)$ is the 4-volume of region encompassed by the past-light cone of point $M$ and the constant time surface with $\tau=\tau_b$. 

To calculate $\Omega\left(\tau_M\right)$, we would like to use the conformal coordinate system, in which
\ba
ds^2=a^2\left(\eta\right)\left[d\eta^2-\left(d\chi^2+\sinh^2\chi d\Omega_2^2\right)\right].
\ea
The conformal time is given by
\ba
\eta&&= \int_0^{\tau_b}\frac{d\tau}{a_1\left(\tau\right)}+ \int_{\tau_b}^{\tau}\frac{d\tau}{a_2\left(\tau\right)} \nonumber\\
&&= \eta_b+\frac{\tau_b}{a_b\left(n-1\right)}\left(1-\left(\frac{\tau_b}{\tau}\right)^{n-1}\right).
\ea
where we defined $\eta_b = \int_0^{\tau_b}\frac{d\tau}{a_1\left(\tau\right)}$. We also define $Z\equiv \eta-\eta_b$, $Z_{\infty} \equiv \frac{\tau_b}{a_b\left(n-1\right)}$ and $Z_M \equiv \eta_M-\eta_b$. 
Then the scale factor can be expressed as
\ba
a_2\left(\eta\right)&&=\left(\frac{\tau_b^n}{a_b\left(n-1\right)^n}\right)^{\frac{1}{n-1}}\left(\frac{\tau_b}{a_b\left(n-1\right)}-\left(\eta-\eta_b\right)\right)^{\frac{n}{1-n}}\nonumber\\
&&\equiv A_2\left(Z_{\infty}-Z\right)^{\frac{n}{1-n}},
\ea
where $A_2 \equiv \left(\frac{\tau_b^n}{a_b\left(n-1\right)^n}\right)^{\frac{1}{n-1}}$. 

The past light cone of point $M$ is bounded by $\chi = \eta_M-\eta$; hence we can write
\ba
\Omega (\tau_M) &&= 4\pi \int_{\eta_b}^{\eta_M}d\eta~ a_2^4\left(\eta\right) \int_0^{\eta_M-\eta} d\chi \sinh^2\chi \nonumber\\
&&=\pi \int_{\eta_b}^{\eta_M}d\eta~ a_2^4\left(\eta\right) \left[2\eta-2\eta_M+\sinh\left(2\eta_M-2\eta\right)\right]\nonumber\\
&&=A_2^4 \pi \int_0^{Z_M}dZ \left(Z_{\infty}-Z\right)^{\frac{4n}{1-n}} \left[2Z-2Z_M+\sinh\left(2Z_M-2Z\right)\right]
\label{B6}
\ea

We are interested in the behavior of $\Omega(\tau_M)$ in the limit $\tau_M \rightarrow \infty$, when  $Z_M$ approaches $Z_{\infty}$ and $a_2$ grows without bound.  In this limit, we can replace $Z_M$  by $Z_\infty$ in the square brackets in (\ref{B6}) and then expand the expression in the square brackets in powers of $(Z_\infty -Z)$.  This gives
%the asymptotic $\Omega$ at large $\tau_M$ is approximated by
\ba
\Omega (\tau_M \to\infty) &&\simeq A_2^4 \pi \int_0^{Z_{M}}dZ \left(Z_{\infty}-Z\right)^{\frac{4n}{1-n}} \left[2Z-2Z_{\infty}+\sinh\left(2Z_{\infty}-2Z\right)\right]\nonumber\\
&&=A_2^4 \pi \sum_{\substack{m = 1} }^{+\infty} \int_0^{Z_{M}}dZ \frac{2^{2m+1}}{\left(2m+1\right)!} \left(Z_{\infty}-Z\right)^{2m+1+\frac{4n}{1-n}} \nonumber\\
&& =A_2^4 \pi \sum_{\substack{m = 1} }^{+\infty} \frac{2^{2m+1}}{\left(\frac{4n}{n-1}-2m-2\right)\left(2m+1\right)!} \left[\left(Z_{\infty}-Z_M\right)^{2m+2+\frac{4n}{1-n}}-Z_{\infty}^{2m+2+\frac{4n}{1-n}}\right]\nonumber\\
&&=a_b^2\tau_b^2\pi  \sum_{\substack{m = 1} }^{+\infty}  \frac{\left(2/\left(n-1\right)\right)^{2m+1}\left(\tau_b/a_b\right)^{2m} }{\left(4n-(2m+2)(n-1)\right)\left(2m+1\right)!} \left[\left(\tau_M/\tau_b\right)^{\left(2-2n\right)m+2n+2}-1\right]\nonumber\\
&&\simeq  \frac{\pi \tau_M^4 }{3\left(n-1\right)^3},
\ea
where we only keep the dominant term, namely the term with $m=1$, in the last line.  Substituting this in Eq. (\ref{PB}), we obtain 
\ba
{\cal P} \left(\tau_M \to\infty \right) \simeq   \frac{4\pi \Gamma  \tau_M^3}{3\left(n-1\right)^3}\exp\left(-\frac{\pi\Gamma }{3\left(n-1\right)^3}\tau_M^4  \right).
\ea

\end{document}